\documentclass[journal=jctc,manuscript=article]{achemso}

\usepackage[version=3]{mhchem} 

\newcommand*{\q}{\mathbf{q}}
\newcommand*{\e}{\mathbf{e}}
\newcommand*{\A}{\mathbf{A}}
\newcommand*{\C}{\mathbf{C}}

\newcommand*{\bb}{\mathbf{b}}
\newcommand*{\OO}{\mathbf{O}}

\newcommand*{\dd}{\mathbf{d}}

\newcommand*{\rN} {\mathbf{r}^{3N}}
\newcommand*{\idrn}{\int \mathrm{d}\rN\,}
\newcommand*{\ebuO}{e^{-\beta U_0(\rN)}}
\newcommand*{\idq} {\int \mathrm{d}\q\,}

\newcommand*{\idke} {\int_{-\infty}^\infty \mathrm{d}k_e\,}
\newcommand*{\idkd} {\int_{-\infty}^\infty \mathrm{d}k_d\,}

\author{Shern R. Tee}
\email{s.tee@uq.edu.au}
\author{Debra J. Searles}
\email{d.bernhardt@uq.edu.au}%
\affiliation{Australian Institute of Bioengineering and Nanotechnology, The University of Queensland, Brisbane, QLD, 4072, Australia}
\alsoaffiliation{School of Chemistry and Molecular Biosciences, The University of Queensland, Brisbane, QLD, 4072, Australia}

\title[Constrained Charge and Constant Potential Simulations]
  {Constant Potential and Constrained Charge Ensembles for Simulations of Conductive Electrodes}

\abbreviations{MD, ConP, ConQ}

\begin{document}

\begin{tocentry}

\includegraphics[width=0.95\textwidth]{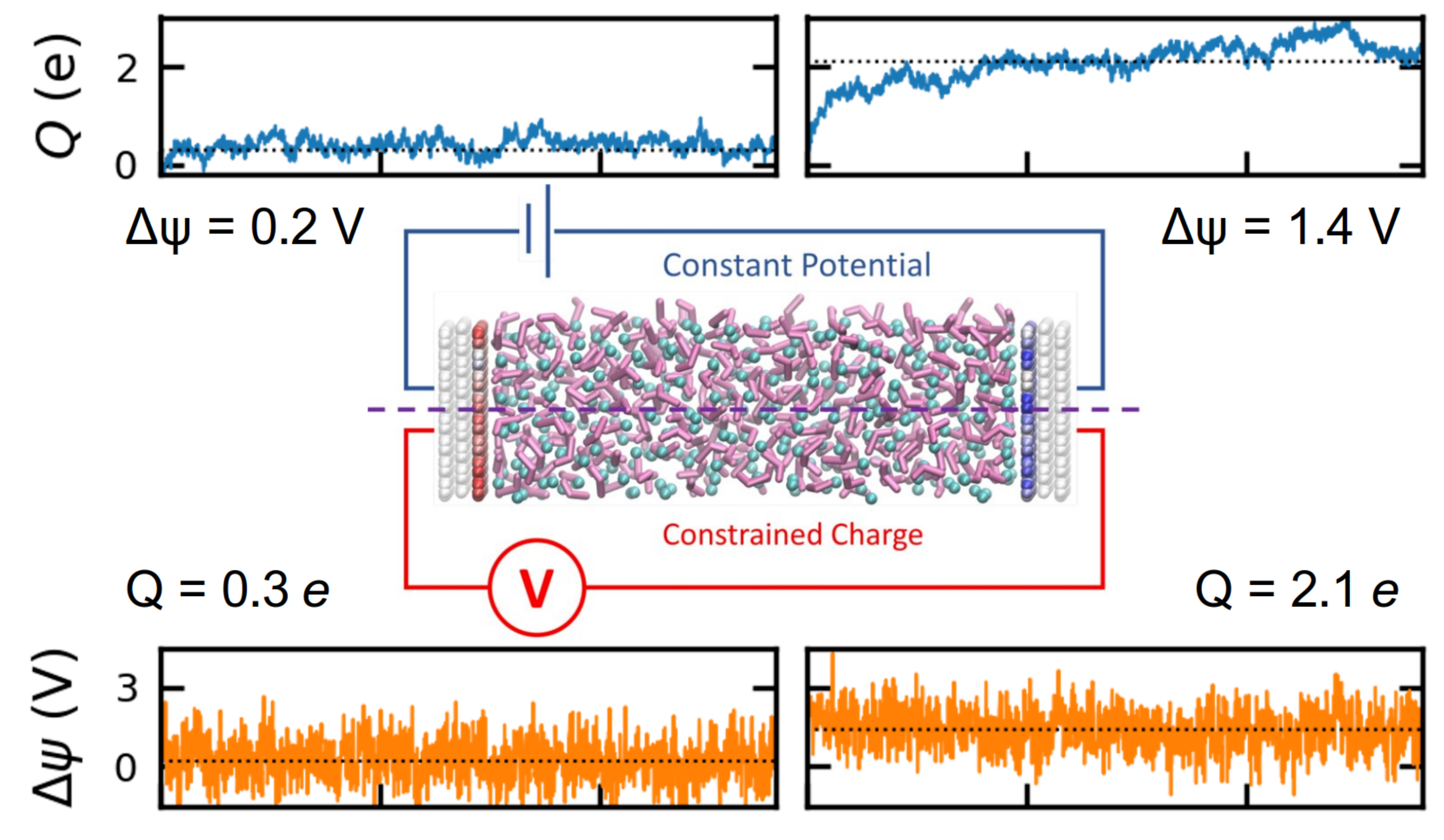}




\end{tocentry}

\begin{abstract}
Constant potential method molecular dynamics simulation (CPM MD) enables the accurate modelling of atomistic electrode charges when studying the electrode-electrolyte interface at the nanoscale. Here we extend the theoretical framework of CPM MD to the case in which the total charge of each conductive electrode is controlled, instead of their potential differences. We show that the resulting thermodynamic ensemble is distinct from that sampled in with a fixed potential difference, but rigorously related as conjugate ensembles. This theoretical correspondence, which we demonstrate using simulations of an ionic liquid supercapacitor, underpins the success of recent studies with fixed total charges on the electrodes. We show that equilibration is usefully sped up in this ensemble and outline some potential applications of these simulations in future.
\end{abstract}

\section{Introduction}

Understanding the interface between a solid electrode and a fluid electrolyte is crucial for making progress in electrochemistry, with wide-ranging applications throughout industry such as in energy storage and catalysis.
While dilute electrolytes are well-treated using current theories of the electric double layer (EDL), more complex situations such as concentrated electrolytes or nano-structured electrodes remain out of reach.
Molecular dynamics (MD) simulations are thus vital for directly modelling and visualizing the electrode-electrolyte interface in more complex situations.
MD simulations using the constant potential method, in particular, enable realistic calculations of the electrode response to the surrounding electrolyte at specific electrode potentials.
In CPM MD, the charges on electrode atoms are dynamically adjusted to ensure each electrode is kept at a pre-determined electric potential.
Thus, these simulations can model dynamic phenomena such as capacitor charging and discharging, and accurately predict the polarization response of electrodes with arbitrary nanostructures.

Recently, two studies\cite{Dufils2021,Zeng2021} have modelled conductive electrodes using similar techniques, but with time-varying constraints on the the total charge on each electrode, instead of the potential difference between them.
These articles coined the terms ``computational amperometry'' and ``galvanostatic mode'' respectively for their technique, in which the electrode charges on a model supercapacitor are increased and then decreased linearly over the course of a simulation.
\citet{Dufils2021} found that ``computational amperometry'' gave better non-equilibrium capacitance measurements compared to analogous ``computational voltmetry'', in which the potential difference was changed instead of the total charge.
\citet{Zeng2021} found that ``galvanostatic mode ConP'' captured experimentally-comparable charging and discharging dynamics.
They further verified that for complex nanoporous electrodes it was vital to distribute charges using a constant potential approach, and not simply place uniform charges on each electrode atom regardless of its position or local environment.

In this paper, we provide a theoretical underpinning for these findings by focusing on the equilibrium distributions of electrode charges and electrolyte configurations, instead of the non-equilibrium aspect of increasing or decreasing these charges over time.
We label the original application of CPM MD `ConP', for constant potential, and label the new total charge technique as `ConQ', for constrained charge.\footnote{We avoid the initialism `CCM' as it has been used by other authors, notably \citet{Zeng2021}, to refer to a uniform distribution charge without any reliance on local potential calculations.}
We show that the ConP and ConQ methods are intimately related, with a physical analogy to placing the simulated supercapacitor in a closed or open electrical circuit respectively.
This helps us analyze the statistical mechanics of ConP and ConQ ensembles, and rationalize both the similarities and differences between ConP and ConQ in terms of a Laplace transform between their partition functions.
Potential difference and electrode charge are thus shown to be an intensive-extensive pair of conjugate thermodynamic variables.
This sheds light on both the thermodynamic equivalence between expectation values the ConP and ConQ ensembles, and the fundamental difference in the magnitude and timescale of fluctuations in each.
We finally discuss ConP and ConQ simulation results of a typical supercapacitor modelled computationally with graphene electrodes and coarse-grained ionic liquid electrolyte, which demonstrate clearly the utility and weaknesses of both approaches.
Two recent papers have described charge-constrained molecular dynamics in terms of chemical potential equalization in an open circuit \cite{Sato2021,Sato2022}, and complement the results in this work about the statistical mechanics of the resulting configurational ensemble.
Thus, both ConP and ConQ simulations represent physically valid descriptions of electrochemical systems, but with different use cases.
A ConP simulation at a given potential difference samples a wider range of possible configurations than a ConQ simulation at a given electrode charge.
The configurational overlap between ConP simulations allows for histogram reweighting techniques to improve statistical efficiency \cite{Limmer2013}.
On the other hand, ConQ simulations appear to have shorter equilibration and correlation times.
Furthermore, they offer the possibility of meaningful comparison with results from quantum chemistry, such as density functional theory (DFT)-based calculations.
Of special relevance, recent DFT calculations have studied electrode-electrolyte interfaces at different potentials by inducing a fixed amount of charge on the electrode, using either explicit counter-charges \cite{AuH2O_explicitcharge_2021} or an ionically-imbalanced electrolyte composition \cite{ILGraphene_DFT_2020, PtH2O_ions_2020, AgH2O_diffions_2022}.
Parallel ConQ simulations for these systems could enable computational chemists to combine the quantitative accuracy of DFT simulations with the larger length and time scales of MD simulations.

\section{Theory}

\subsection{ConP Charges and Energies}

\begin{figure}
    \centering
    \includegraphics[width=0.6\textwidth]{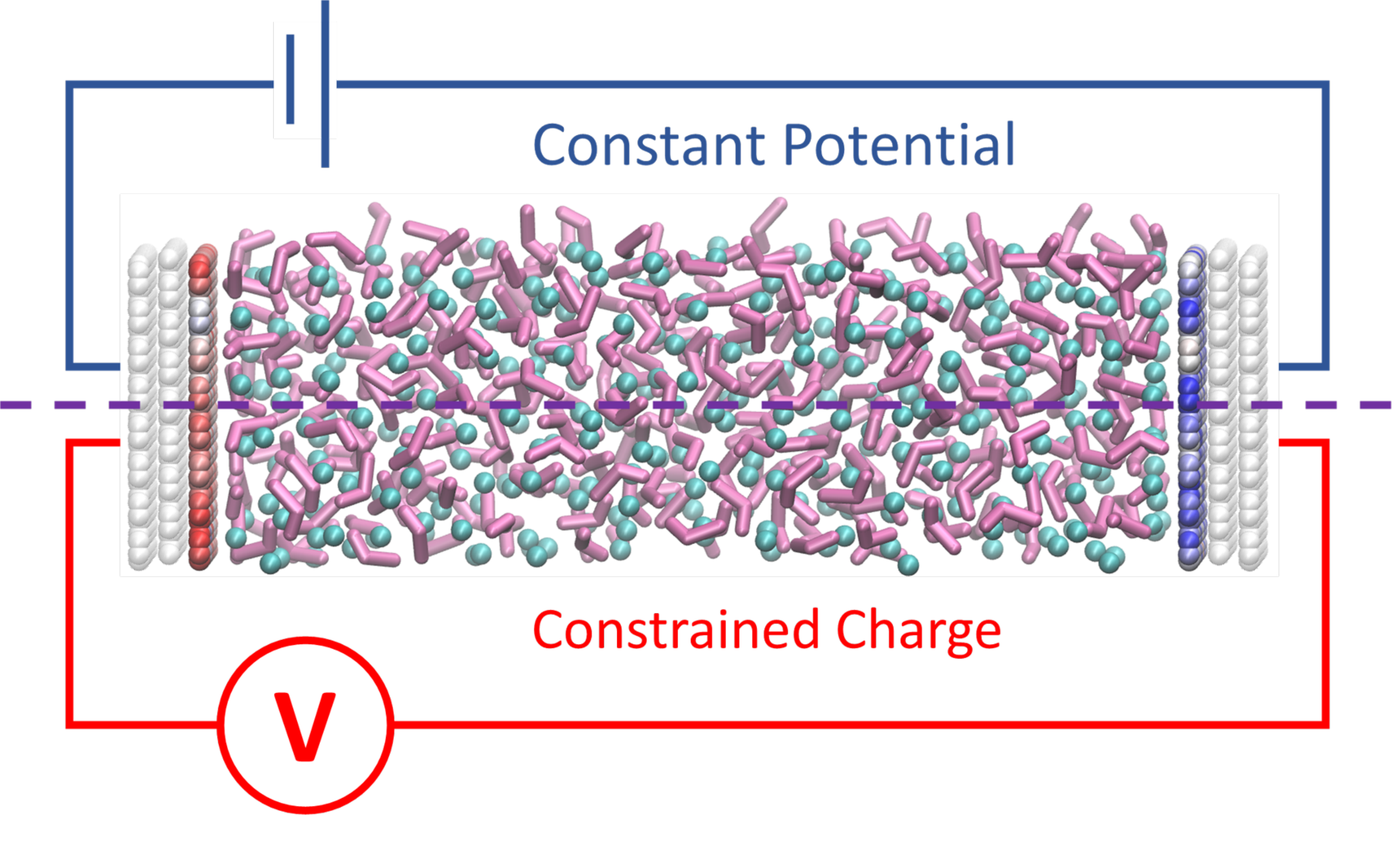}
    \includegraphics[width=0.6\textwidth]{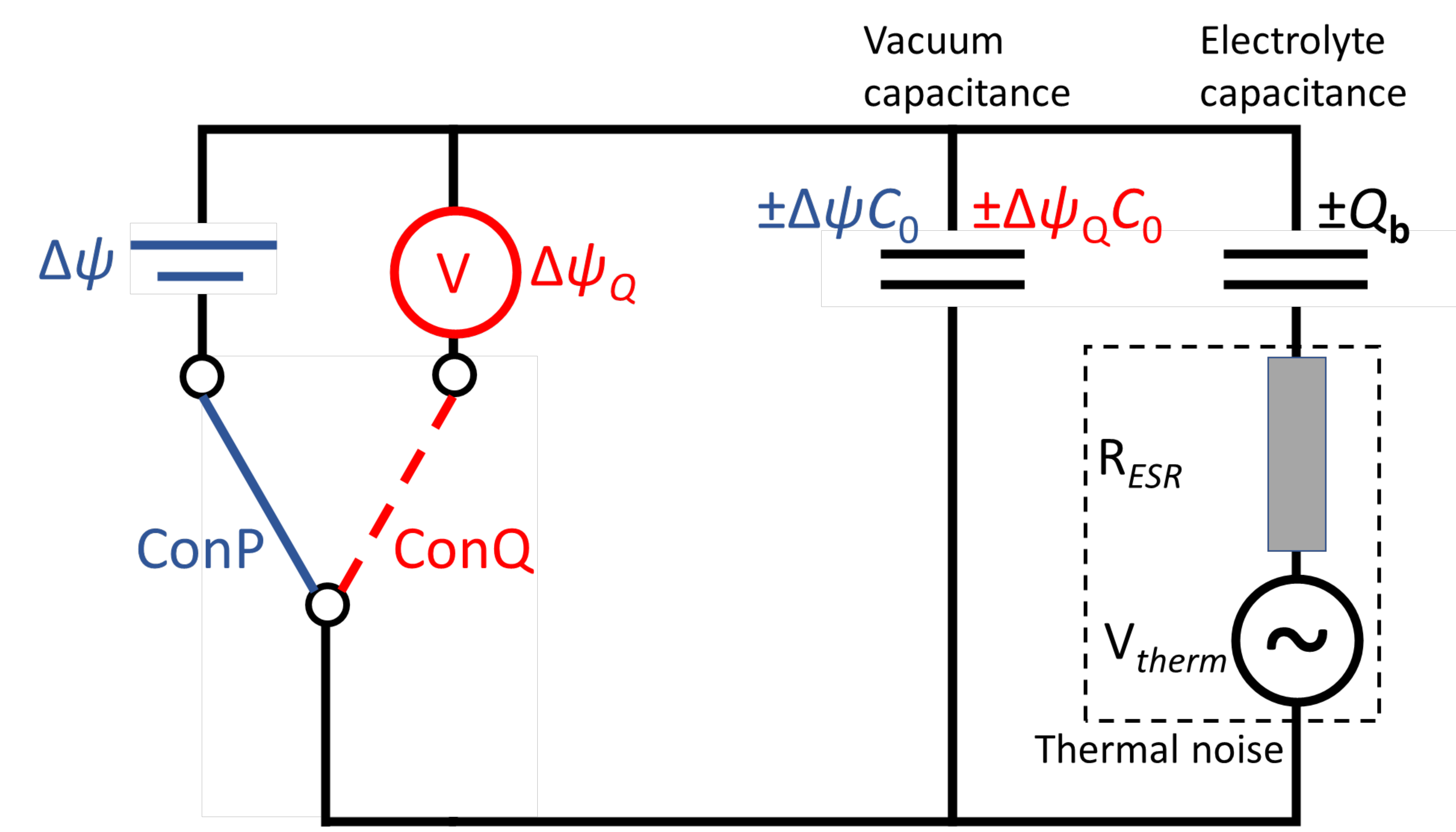}
    \caption{Simulation system and conceptual circuit for comparing ConP and ConQ ensembles. (a) A typical simulated supercapacitor consists of an electrolyte (magenta and cyan) confined between two electrodes (white). Induced charges are dynamically updated on the electrodes dependent on the electrolyte configuration, depicted with red (blue) shades on the negative (positive) electrode. The electrodes can hold either (top) ConP charges, with a fixed potential difference, or (bottom) ConQ charges, with a fixed total charge. (b) The simulated supercapacitor is equivalent to a vacuum capacitor with charge $\pm \Delta \psi$ $C_0$ (center) in parallel with the electrolyte response capacitor with instantaneous charge $\pm Q_{\mathbf{b}}$ (right). In ConP simulations (blue), potential difference $\Delta \psi$ is fixed and the total charge $Q_P = Q_\mathbf{b} + \Delta \psi C_0$ fluctuates; in ConQ simulations (red), total charge $Q$ is fixed and the potential difference $\Delta \psi_Q = (Q - Q_\mathbf{b})/C_0$ fluctuates. In both cases fluctuations are due to the Johnson-Nyquist noise of the effective series resistance associated with the electrolyte; the vacuum capacitance has no associated noise due to the Born-Oppenheimer procedure \cite{Scalfi2020ChargeEnsemble}.}
    \label{fig:overview_circuit}
\end{figure}

We briefly recap the theory behind ConP constant potential molecular dynamics simulations, before extending the theory to ConQ simulations.
An electrolyte fluid in the presence of electrified electrodes is considered (see Figure \ref{fig:overview_circuit} (a)).
The electrode-electrolyte configuration consists of the positions of the $N$ electrolyte particles, $\mathbf{r}^{3N}$, and the charges of the $M$ electrode particles, $\mathbf{q}$.
(Unless otherwise specified, vectors like $\mathbf{q}$ are $M$-dimensional).
We assume that the electrolyte particles have fixed charge and the electrode particles have fixed positions.
%

We can separate the electrode-independent and electrode-dependent terms of the inter-particle potential energy, $U$:
\begin{equation}
    U(\mathbf{r}^{3N}, \mathbf{q}) = U_0(\mathbf{r}^{3N}) + \frac{1}{2} \mathbf{q}^T \mathbf{A} \mathbf{q} - \mathbf{b}^T(\mathbf{r}^{3N}) \mathbf{q},
    \label{eqn:def_U}
\end{equation}
omitting any potential difference or charge constraints for now.
The first term, $U_0(\mathbf{r}^{3N})$, is the electrode-independent potential energy, and covers all non-Coulombic interactions and all electrolyte-electrolyte Coulombic interactions between electrolyte particles.
The second and third terms cover the Coulombic interactions of the electrode charges with other electrode charges and electrolyte charges respectively.
The matrix $\mathbf{A}$ in the second term represents the mutual capacitances between electrode particles, while the vector $\mathbf{b}(\mathbf{r}^{3N})$ in the third term represents the electrostatic potential experienced by electrode charges due to electrolyte particles and depends on their positions.

If we now connect the electrodes to charge reservoirs separated by some potential difference $\Delta \psi$, the total potential energy becomes 
\begin{equation}
U_P(\mathbf{r}^{3N}, \mathbf{q}, \Delta \psi)=U(\mathbf{r}^{3N}, \mathbf{q})- \Delta \psi \mathbf{d}^T \mathbf{q}
\label{eqn:def_U_P}
\end{equation}
where the vector $\mathbf{d}$ is an ``indicator'' vector, denoting which electrode particle belongs to which electrode.
That is, if there are $n_L$ ($n_R$) particles in the left (right) electrode, $\mathbf{d}$ is the $M$-entry vector
\begin{equation}
    \mathbf{d} \equiv \{ \alpha_L, \cdots, \alpha_L, \alpha_R, \cdots, \alpha_R \}
    \label{eqn:def_d}
\end{equation}
where the first $n_L$ entries have value $\alpha_L = n_R / (n_L + n_R)$ and the subsequent $n_R$ entries have value $\alpha_R = \alpha_L - 1$.
(For symmetric electrodes $n_L = n_R$, and therefore $\alpha_L = 1/2$, $\alpha_R = -1/2$.)
%
%
We refer the reader to other papers for further details, including explicit expressions for each term \cite{Wang2014EvaluationCapacitors,Scalfi2020ChargeEnsemble,Tee2022}.

In ConP simulations, at each time step the electrode charges $\mathbf{q}$ are updated to fix the potential difference between the electrodes, and this minimizes the total potential energy, $U_P$.
This contrasts with conventional MD, in which $\mathbf{q}$ would be constant.
The Born-Oppenheimer-style charge update procedure reflects how, in reality, charge redistribution occurs instantaneously on the time scale of an MD time step \cite{Reed2007ElectrochemicalElectrode,Merlet2013SimulatingSurfaces}.
In ConP simulations, the charge $\mathbf{q}_P$ is
\begin{equation}
    \mathbf{q}_P(\Delta \psi, \mathbf{b}) = \mathbf{O}\mathbf{C} (\mathbf{b} + \Delta \psi \mathbf{d}),
    \label{eqn:soln_qstar}
\end{equation}
where $\mathbf{C} \equiv \mathbf{A}^{-1}$ is the capacitance matrix, and the electroneutrality projection matrix $\mathbf{O}$ keeps the total charge of the system at zero.
We include an expression for $\mathbf{O}$ in Appendix \ref{sec:SI_fullzq} and refer the reader to references \citenum{Scalfi2020ChargeEnsemble} and \citenum{Tee2022} for further details.
(`P' and `Q' subscripts will denote ConP and ConQ versions, respectively, of identical or closely-related physical quantities.)
The instantaneous total charge on one electrode is then
\begin{equation}
    Q_P(\Delta \psi, \mathbf{b}) = \mathbf{d}^T\mathbf{q}_P(\Delta \psi, \mathbf{b}) = \mathbf{d}^T\mathbf{O}\mathbf{C} (\mathbf{b} + \Delta \psi \mathbf{d}) \equiv Q_\mathbf{b} + \Delta \psi C_0.
\end{equation}
To maintain electroneutrality, the charge on the other electrode will be $-Q_P$.
The quantities $C_0 \equiv \mathbf{d}^T \mathbf{OCd}$ and $Q_\mathbf{b} \equiv \mathbf{d}^T \mathbf{OCb}$ have physical significance (Figure \ref{fig:overview_circuit}(b)).
Without any electrolyte (i.e. $\mathbf{b}=\mathbf{0})$, the electrodes would have a ``vacuum capacitance''\cite{Scalfi2020ChargeEnsemble} $C_0$, developing a charge $\pm \Delta \psi C_0$ at potential difference $\Delta \psi$.
With electrolyte present, the electrolyte's dielectric response further polarizes the electrodes and induces an additional electrode charge $\pm Q_{\mathbf{b}}$.
In ConP simulations, the electrodes display realistic charging and discharging behaviour, and settle into an equilibrium distribution with thermally fluctuating charges.
We can interpret this physically as the ``electrolyte capacitance'' having a non-zero equivalent series resistance (ESR).
The ESR in turn shows Johnson-Nyquist noise at a finite temperature, represented by an added thermal potential difference which fluctuates about zero mean.

Figure \ref{fig:overview_circuit}(b) shows the final physical picture of a simulated supercapacitor in ConP ensemble.
Figure 1(b) shows the final physical picture of a simulated supercapacitor in ConP ensemble.
Circuit theory could give analytical relationships between the thermal noise, modelled by the electrolyte ESR, and the (ConP) electrode charge or (ConQ) potential difference noise, but we do not pursue that line of thought here.

Figure \ref{fig:overview_circuit}(b) shows the final physical picture of a simulated supercapacitor in ConP ensemble.
Circuit theory could give analytical relationships between the thermal noise, modelled by the electrolyte ESR, and the (ConP) electrode charge or (ConQ) potential difference noise, but we do not pursue that line of thought here.
Interestingly, a recent preprint describes how the frequency-dependent impedance of nanocapacitors can be calculated from the charge fluctuations in a ConP simulation \cite{Pireddu}.
The physical picture here ties in with that approach, since the thermal non-ideality of the electrolyte configuration is both theoretically the source of a nanocapacitor's impedance and practically the source of electrode charge fluctuations in a ConP simulation.

\subsection{ConQ Charges and Energies}

In ConP simulations, we specify the potential difference between electrodes and observe the electrode charges in response.
In ConQ simulations, we no longer specify the potential difference, but we constrain the total electrode charges to be $\pm Q$, i.e. $\mathbf{d}^T \mathbf{q}_Q = Q$. Meanwhile, as in ConP simulations, the conditions of charge neutrality and equal electrostatic potential at each atom of an electrode are maintained.
Then the Born-Oppenheimer charge, $\mathbf{q}_Q$, is
\begin{equation}
    \mathbf{q}_Q(Q, \mathbf{b}) = \mathbf{O}\mathbf{C} (\mathbf{b} + \Delta\psi_Q (Q, \mathbf{b})  \mathbf{d})
    \label{eqn:soln_qdagger}
\end{equation}
with the instantaneous potential difference $\Delta \psi_Q$
\begin{equation}
    \Delta \psi_Q(Q, \mathbf{b}) = \frac{Q - \mathbf{d}^T \mathbf{O}\mathbf{C}\mathbf{b}}{\mathbf{d}^T \mathbf{O}\mathbf{C}\mathbf{d}} \equiv \frac{Q - Q_\mathbf{b}}{C_0}.
    \label{eqn:soln_deltapsidagger}
\end{equation}
We can obtain some physical intuition about this expression if we imagine flipping the switch in Figure \ref{fig:overview_circuit}(b) from the blue branch (solid) to the red branch (dashed).
Suppose we specify a set of charges over the vacuum capacitor, $Q_0 = \Delta \psi C_0$, and the ``electrolyte capacitor'' $Q_\mathbf{b}$.
Our instantaneous specification will be identical whether the circuit switch connects the capacitors to a potential difference, as in a ConP simulation, or to a voltmeter, as in a ConQ simulation (resulting in open circuit conditions).
Since (ideally) no current flows through a voltmeter, the total charge of the vacuum and electrolyte capacitors is now constant, but now their distribution between capacitors fluctuates with the thermal fluctuations of the ESR.
Thus the potential difference \emph{measured} by the voltmeter also fluctuates in a ConQ simulation -- as opposed to being imposed as a constant value in a ConP simulation -- and, at any one time, the measured potential difference will be given precisely by equation \eqref{eqn:soln_deltapsidagger}.
Put differently, a given configuration is equally valid for a ConP simulation at potential difference $\Delta \psi$ as for a ConQ simulation at total charge $Q$.
But the configuration may certainly be more, or less, \emph{probable} given any one simulation protocol.

The physical equivalence, just described, between ConP and ConQ simulations certainly suggests that they are thermodynamically equivalent ensembles -- but we need to prove this formally using statistical mechanics.
To do so, we first write down the different energies that a given configuration will have in ConP and ConQ simulation.
The ConP energy comes from substituting the ConP charges $\mathbf{q}_P$ into the potential-dependent $U_P$ from equation \eqref{eqn:def_U_P}, while the ConQ energy comes from substituting the ConQ charges $\mathbf{q}_Q$ into the potential energy $U$ from equation \eqref{eqn:def_U}.
Then the ConP configurational energy $U_P(\mathbf{r}^{3N},\Delta \psi)$ and ConQ configurational energy $U_Q(\mathbf{r}^{3N}, Q)$ are:
\begin{align}
    U_P(\mathbf{r}^{3N},\Delta \psi) &= U_0 - \frac{1}{2} \mathbf{b}^T \mathbf{OCb} - \frac{1}{2} \Delta \psi^2 C_0 - \Delta \psi Q_\mathbf{b} 
    \label{eqn:soln_Ustar_abs}\\
    &\equiv U_0 - U_\mathbf{b} + \frac{1}{2}C_0 \Delta \psi^2  - \Delta \psi Q_P(\Delta \psi,\mathbf{b})
    \label{eqn:soln_Ustar_rel}\\
    U_Q(\mathbf{r}^{3N},Q) &=U(\mathbf{r}^{3N},Q) = U_0 - U_\mathbf{b} + \frac{1}{2}C_0 \Delta \psi_Q(Q,\mathbf{b})^2
    \label{eqn:soln_Udagger}
\end{align}
In either the ConP or the ConQ case, the electrode charges $\mathbf{q}$ are fully dependent on the electrolyte configuration, and are no longer independent degrees of freedom.

Again, the terms in each energy expression are physically intuitive.
Both energies sum together the electrode-independent energy $U_0$, the `electrolyte offset' $U_{\mathbf{b}} \equiv \frac{1}{2} \mathbf{b}^T\mathbf{OCb}$, and the vacuum charging energy $\frac{1}{2} C_0\Delta \psi^2$ (with $\Delta \psi$ being a parameter in ConP simulations and a phase variable in ConQ simulations).
The additional term $-\Delta \psi Q_P$ in the ConP energy $U_P$ (equation \eqref{eqn:soln_Ustar_rel}) is the work of drawing from charge reservoirs at potential difference $\Delta \psi$ to charge the capacitor plates.
(The negative sign for $U_{\mathbf{b}}$ reflects how the electrode induced charges stabilize the polarization of the electrolyte, as the potential difference across the supercapacitor increases.
Recall that $U_0$ includes all intra-electrolyte Coulombic interactions.
Thus, as the magnitude of $\mathbf{b}$ increases, both $U_0$ on its own and the combination $U_0 - U_{\mathbf{b}}$ will increase.)
%
%

Comparing $U_P$ in equation \eqref{eqn:soln_Ustar_rel} and $U_Q$ in equation \eqref{eqn:soln_Udagger} shows that they are related by a Legendre transform, differing by the term $\Delta \psi Q$.
Thus, ConP and ConQ simulations are connected by the thermodynamic relationship between potential difference and total electrode charge as an intensive-extensive pair of conjugate variables.

We now show that, as expected, the ConP and ConQ partition functions are related by a Laplace transform, and thus the ConP and ConQ statistical ensembles are thermodynamically equivalent.

\subsection{Transforming between ConP and ConQ ensembles}

We can write down the ConP configurational partition function $Z_P$, 
\begin{equation}
    Z_P(\Delta \psi) \equiv \int e^{-\beta U_P(\mathbf{r}^{3N},\Delta \psi)} \, \mathrm{d}\mathbf{r}^{3N} \label{eqn:def_zp}
\end{equation}
and the ConQ configurational partition function,
\begin{equation}
    Z_Q(Q) \equiv \int e^{-\beta U_Q(\mathbf{r}^{3N},Q)} \, \mathrm{d}\mathbf{r}^{3N}.
\end{equation}

The main novel result of our paper is that these partition functions are related by a (scaled) Laplace transform:
\begin{equation}
    Z_P(\Delta \psi) = \frac{1}{\sqrt{2 \pi k_B T C_0}} \int_{-\infty}^{\infty} e^{\beta Q \Delta \psi} Z_Q (Q)\,\mathrm{d}Q. \label{eqn:soln_legendre}
\end{equation}

To verify this mathematically, we first write out the integral on the right hand side of equation \eqref{eqn:soln_legendre} and interchange the order of integration (between $Q$ and $\mathbf{r}^{3N}$).
Rearranging the exponential energy terms, and substituting the definition of $U_P(\Delta \psi, \mathbf{b})$ from equation \eqref{eqn:soln_Ustar_abs}, then gives:
\begin{align}
    \int_{-\infty}^{\infty} e^{\beta Q \Delta \psi} &Z_Q(Q)\, \mathrm{d}Q = \int \mathrm{d}\mathbf{r}^{3N} \, \bigg[ e^{-\beta U_P(\mathbf{r}^{3N}, \Delta \psi)} \nonumber \\  &\int_{-\infty}^{\infty}\mathrm{d}Q\, e^{-\frac{\beta}{2 C_0}(Q - Q_\mathbf{b}-\Delta \psi C_0)^2}\bigg]. \label{step:soln_legendre}
\end{align}
Since equation \eqref{step:soln_legendre} is just $Z_P$ from equation \eqref{eqn:def_zp} multiplied by a constant Gaussian integral, we thus have equation \eqref{eqn:soln_legendre}.

Equation \eqref{eqn:soln_legendre} was \emph{derived} by considering the ``full'' partition functions for the ConP and ConQ ensembles, in which $\mathbf{q}$ is allowed to freely vary.
This was inspired by work in reference [\citet{Scalfi2020ChargeEnsemble}], where it was shown that using Born-Oppenheimer dynamics causes reduced variance of some observables in the Born-Oppenheimer ConP ensemble compared to the full constant potential ensemble.
We leave this derivation in the Appendix for brevity, and simply mention the significance of the scaling factor $1/\sqrt{2 \pi k_B T C_0}$.
Both Born-Oppenheimer partition functions $Z_Q(Q)$ and $Z_P(\Delta \psi)$ are integrals over $3N$ degrees of freedom (of the electrolyte positions $\mathbf{r}^{3N}$).
Therefore, $Z_P$ is one degree-of-freedom short of allowing an exact Laplace transform to $Z_Q$.
However, the full ConP ensemble does indeed have one more degree of freedom than the full ConQ ensemble, and the Laplace transform between them has no additional scaling factor, as the derivation in the Appendix shows.
In any case, since this scaling factor is configuration-independent, it is not relevant to the subsequent thermodynamic comparisons between ensembles.
%
%

%

\subsection{Thermodynamic Equivalence: Correspondence of ConP and ConQ Averages}

We can now explore thermodynamic relationships between ConP and ConQ configurational ensembles using familiar statistical mechanical tools.
We start by defining the ConQ free energy:
\begin{equation}
    F_Q(Q) \equiv - k_B T \ln Z_Q(Q).
    \label{eqn:def_FQ}
\end{equation}
As usual, the derivative $\partial F_Q(Q)/\partial Q = \langle \partial U_Q /\partial Q \rangle$ is the average potential difference:
\begin{equation}
    \langle \Delta \psi_Q\rangle(Q) =  \frac{\int \Delta \psi_Q(\mathbf{r}^{3N},Q)\, e^{-\beta U_Q(\mathbf{r}^{3N},Q)} \, \mathrm{d}\mathbf{r}^{3N}}{\int e^{-\beta U_Q(\mathbf{r}^{3N},Q)} \, \mathrm{d}\mathbf{r}^{3N}} = \frac{\partial F_Q(Q)}{\partial Q}.
    \label{eqn:soln_dFdQ}
\end{equation}
Differentiating again gives an expression for (inverse) differential capacitance, which we label $C^D_Q$ for now:
\begin{equation}
    C^D_Q(Q) \equiv \left(\frac{\partial^2 F_Q(Q)}{\partial Q^2} \right)^{-1}
    \label{eqn:def_CdaggerD}
\end{equation}
We will write and analyse an explicit expression for $C^D_Q$ when discussing fluctuations in the ConQ ensemble.

%
%
Now, categorizing electrolyte configurations by the induced ConP charge $Q_P(\mathbf{r}^{3N}, \Delta \psi)$, we have
\begin{equation}
    Z_P(\Delta \psi) = \int_{-\infty}^{\infty} \left[ \int e^{-\beta U_P(\mathbf{r}^{3N},\Delta \psi)} \delta(Q_P(\mathbf{r}^{3N}, \Delta \psi) - Q) \, \, \mathrm{d}\mathbf{r}^{3N} \right] \mathrm{d}Q
    \label{eqn:step_probdistQ}
\end{equation}
Comparing integrands with equation \eqref{eqn:soln_legendre}, and using the definition of a probability distribution, we have
\begin{equation}
    \mathrm{Pr}(Q_P(\mathbf{r}^{3N},\Delta \psi) = Q) \propto e^{\beta [Q \Delta \psi - F_Q(Q)]}
    \label{eqn:soln_probdistQ}
\end{equation}
(up to a normalization constant).

Now in the thermodynamic limit for the ConP ensemble the probability distribution of $Q_P$ will have a single sharp peak.
We can thus approximate the probability distribution in equation \eqref{eqn:soln_probdistQ} as Gaussian by linearizing the log-probability about its peak:
\begin{equation}
    Q\Delta \psi - F_Q(Q) \approx -F_Q(Q_0) - \frac{1}{2} \frac{(Q - Q_0)^2}{C^D_Q(Q_0)},
\end{equation}
where the mean charge $Q_0$ obeys the condition
\begin{equation}
    \langle \Delta \psi_Q \rangle (Q_0) = \left. \frac{\partial F_Q(Q)}{\partial Q}\right|_{Q_0} = \Delta \psi.
\end{equation}
Now, clearly $Q_0$ will also be the average charge in the ConP ensemble at potential difference $\Delta \psi$ -- that is, $\langle Q_P \rangle [\Delta \psi] = Q_0$.
But then we have shown that
\begin{equation}
    \langle Q_P \rangle [\langle \Delta \psi_Q \rangle (Q_0)] = Q_0.
\end{equation}
That is, taking the equilibrium average potential difference of a ConQ ensemble at specified charge is the inverse of taking the average charge of a ConP ensemble.
The capacitance of a capacitor will therefore be identical when calculated between ConP and ConQ ensembles, given our earlier assumption of a sharply-peaked distribution in the induced electrode charge.

This relationship only applies wherever the ConP ensemble has a sharply-peaked distribution in the induced electrode charge.
On the other hand, prior ConP simulation studies of supercapacitors have observed sharp jumps of induced charge at specific potential differences, corresponding to transitions in the electrolyte packing near electrodes \cite{Merlet2014EDLLife}.
The current picture allows us to be rigorous about the ``phase transitions'' entailed in these capacitance jumps:
we expect these jumps to involve ranges of the electrode charge $Q$ over which the free energy $F_Q (Q)$ changes little.
Furthermore, we expect that the ConQ ensemble will allow fine-grained explorations of the phase space near those phase points, where ConP studies are bogged down by the long time scale of phase transitions.

In addition, the ConQ ensemble directly corresponds to quantum-mechanical DFT calculations of charged electrode-electrolyte interfaces.
These calculations typically involve a fixed amount of charge induced on the electrode, as mentioned in the Introduction.
By contrast, grand-canonical DFT approaches with constant electrode potential and varying electrode charge result in a non-neutral simulation cell, requiring various correction approaches to avoid a divergent electrostatic energy \cite{GCDFT}.
Thus, future work comparing MD and DFT studies of the same system can benefit from direct comparisons of ConQ simulations, with the reassurance of ensemble equivalence with ConP simulations in the thermodynamic limit.

\subsection{Statistical Difference: Variance Trends in the ConQ and ConP Ensembles}

Thermodynamically-equivalent ensembles will show different fluctuations in observable quantities, even in the thermodynamic limit where the observables' averages are identical.
As a simple example, the total energy fluctuates in the canonical ensemble (and is related to the specific heat capacity), but does not fluctuate in the microcanonical ensemble, and fluctuations of other quantities like kinetic energy will similarly differ \cite{Lebowitz1967}.
In this section we show how fluctuations differ for ConP and ConQ ensembles.
In particular, we show that the ConQ ensemble has a much narrower distribution of electrolyte configurations, making its dynamics much faster than the equivalent ConP ensemble.
This has been observed in other recent studies \cite{Dufils2021,Zeng2021}.
However, as far as we know, this is the first time such observations have been explained from a rigorous statistical mechanical standpoint.

We start with the variance of the observed potential difference in the ConQ ensemble, $\langle \delta (\Delta \psi_Q)^2 \rangle$.
Starting from equation \eqref{eqn:soln_dFdQ}, we can evaluate the second derivative of $F_Q(Q)$ with respect to $Q$.
Rearranging for the variance gives:
\begin{equation}
    \beta \left\langle \delta(\Delta \psi_Q)^2 \right\rangle (Q) = 1/C_0 - 1/C^D_Q,
    \label{eqn:soln_varpd}
\end{equation}
where $C^D_Q$ is the inverse second derivative of $F_Q(Q)$ (equation \eqref{eqn:def_CdaggerD}).
Two consequences of this equation stand out.
Firstly, since larger systems generally have larger capacitances, the variance in observed potential difference will \emph{decrease} with increasing system size.
Again, this parallels how, for example, simulating systems with larger volumes results in decreased fluctuations in pressure and is consistent with $\Delta \psi$ being intensive whereas $Q$ is extensive.

Secondly, in systems of interest (such as supercapacitors) the electrolyte will contribute significantly to capacitance, and thus $C^D_Q$ will usually be much larger than $C_0$ (the capacitance in the absence of an electrolyte).
Thus, in the ConQ ensemble, the variance in potential difference is dominated by the vacuum contribution, $1/C_0$, and calculating the overall capacitance $C^D_Q$ from the variance is impractical.
For example, in a typical system where $C^D_Q$ is ten times larger than $C_0$, the variance would need to be determined with an accuracy of 1\% to determine $C^D_Q$ with an accuracy of 10\% using equation \eqref{eqn:soln_varpd}.

The determination of variance also explains why dynamics can be significantly accelerated in the ConQ ensemble relative to the ConP ensemble.
We can calculate the variance of the electrolyte-induced charge $Q_\mathbf{b}$ in both ensembles to put them on equal footing.
In the ConQ ensemble, since $Q_\mathbf{b} = Q - C_0 \Delta \psi_Q$, the variance of $Q_\mathbf{b}$ is
\begin{equation}
    \beta \langle \delta Q_\mathbf{b}^2 \rangle_Q = C_0\left(1 - \frac{C_0}{C^D_Q}\right) = \frac{C_0}{C^D_Q} \left(C^D_Q - C_0 \right).
    \label{eq:varconq}
\end{equation}
For comparison, in the ConP ensemble, we have $Q_\mathbf{b} = Q_P - C_0 \Delta \psi$ and $\beta \langle \delta (Q^{*})^2 \rangle = C^D_P - C_0$,  where $C^D_P$ is the differential capacitance in the ConP ensemble\cite{Scalfi2020ChargeEnsemble}.
Then the variance of $Q_\mathbf{b}$ in the ConP ensemble is
\begin{equation}
    \beta \langle \delta Q_\mathbf{b}^2 \rangle_P = C^D_P - C_0.
    \label{eq:varconp}
\end{equation}
$C_D$ is usually much larger than $C_0$, and thus in the ConP ensemble $Q_\mathbf{b}$ (and more generally the electrolyte configuration) will show much larger fluctuations than in the ConQ ensemble.
Indeed, assume that the ensembles are thermodynamically equivalent so that both differential capacitances are approximately equal: $C^D_P \approx C^D_Q \equiv C_D$. Then we have
\begin{equation}
    \frac{\langle \delta Q_\mathbf{b}^2 \rangle_P}{\langle \delta Q_\mathbf{b}^2 \rangle_Q} \approx \frac{C_D}{C_0}.
\end{equation}
The much smaller variance of $Q_\mathbf{b}$ in the ConQ ensemble explains why ConQ simulations tend to result in much faster equilibration times.
As computational results show, in ConQ simulations, long-time correlations can be reduced and thus a simulation of the same duration can have higher statistical efficiency.

On the other hand, the reduced variance of $Q_\mathbf{b}$ in ConQ simulations implies that there is little configurational overlap between simulations at different imposed electrode charges, compared to the overlap in similar ConP simulations.
The configurational overlap in ConP simulations enables the use of histogram reweighting techniques, where the variation of an observable can be interpolated continuously from simulations at different discrete values of the imposed potential difference.
By contrast, histogram reweighting is unlikely to be efficient for ConQ simulations, and so we do not try to develop the theory for how ensembles at different imposed electrode charges are related.
However, the reduced configurational variance and correlation times observed in ConQ simulations might be beneficial for non-equilibrium contexts.
Notably, two prior papers which have used ConQ simulations have focused immediately on `amperometric' applications, in analogy with experimental approaches.
The theory developed here explains why such computational approaches are fruitful, but also suggests that techniques like non-equilibrium umbrella sampling \cite{doi:10.1063/1.3680601} can be fruitfully applied in ConQ simulations to quickly scan the response of the electrode-electrolyte interface to a wide range of applied potentials.
This is outside the scope of our current paper but merits further investigation.

\section{Simulations}

We apply ConP and ConQ simulations to a computational ionic liquid supercapacitor, depicted in Figure \ref{fig:overview_circuit}(a).
The supercapacitor electrolyte consists of 320 ion pairs of coarse-grained 1-butyl-3-methylimidazolium hexafluorophosphate (BMim$^+$-PF$_6^-$), with cations modelled as rigid three-site molecules and anions modelled as single particles.
The electrolyte was confined in a 109.75 \AA{} gap between two electrodes of size 32.2 $\times$ 34.4 \AA{}$^2$, each composed of three layers of atomistic graphene.
Non-Coulombic interactions were modelled as Lennard-Jones (12-6) potentials with a 16 \AA{} cutoff, and particle-particle-particle-mesh (PPPM) gridding was used to evaluate long-ranged electrostatic interactions with a relative accuracy of $10^{-7}$.
The system was periodic in $x$ and $y$ directions, parallel to the electrodes, while a slab correction of 3$\times$ was used to enforce non-periodicity in the $z$ direction.

The supercapacitor was simulated at a temperature of 400K by applying a Nos\'{e}-Hoover thermostat to the electrolyte with a time constant of 100 fs, with an integration timestep of 2 fs.
The electrode atoms were held immobile.
Constant potential simulations were carried out by updating the charges on the first layer closest to the electrolyte of each electrode every timestep, with grid-based long-range electrostatic calculations \cite{Ludwig2021}.
Three initial configurations were obtained from the endpoints of 30-ns simulations run at a constant potential difference of 0.0 V.
From each initial configuration, the system was simulated under ConP potential differences of 0.0 V to 1.5 V (in 0.1 V increments), as well as ConQ electrode charges of 0.0 e (electronic charge) to 2.4 e (in 0.1e increments).
At each condition the system was simulated for 100 ns, with the last 90 ns being used as the production portion.
Using three different initial configurations resulted in three independent sets of trajectories to allow for analysis of statistical error.

We refer readers to an earlier paper which showed computational acceleration in fully periodic ConP (referred to as CPM MD) simulations \cite{Tee2022}, using the same simulation system,  for further technical details.
The LAMMPS \cite{LAMMPS} molecular dynamics software was used to run simulations, with the ELECTRODE \cite{ELECTRODE} package used for ConP and ConQ charge updating.
We found that ELECTRODE charges were statistically similar to charges from the USER-CONP2 package in the previous study at the same potential differences.
We report the total electrode charge $Q$, consistent with our theoretical derivations, in units of electron charge (e).
Thus, we report capacitances in electrons per volt (e V$^{-1}$).
However, often other studies (including our own previous study) report surface charge densities in microcoulombs per square centimeter ($\mu$C cm$^{-2}$) and areal capacitances in microfarads per square centimeter ($\mu$F cm$^{-2}$) instead.
For ease of comparison we note that, based on the unit cell dimensions, 1.00 e of charge equals 1.45 $\mu$C cm$^{-2}$ of surface charge density, and 1.00 e V$^{-1}$ of capacitance in these simulations equals 1.45 $\mu$F cm$^{-2}$ of areal capacitance.

\section{Results}

\subsection{Fluctuations and Averages of Potential Difference and Electrode Charge}

\begin{figure}
    \centering
    \includegraphics[width=\columnwidth]{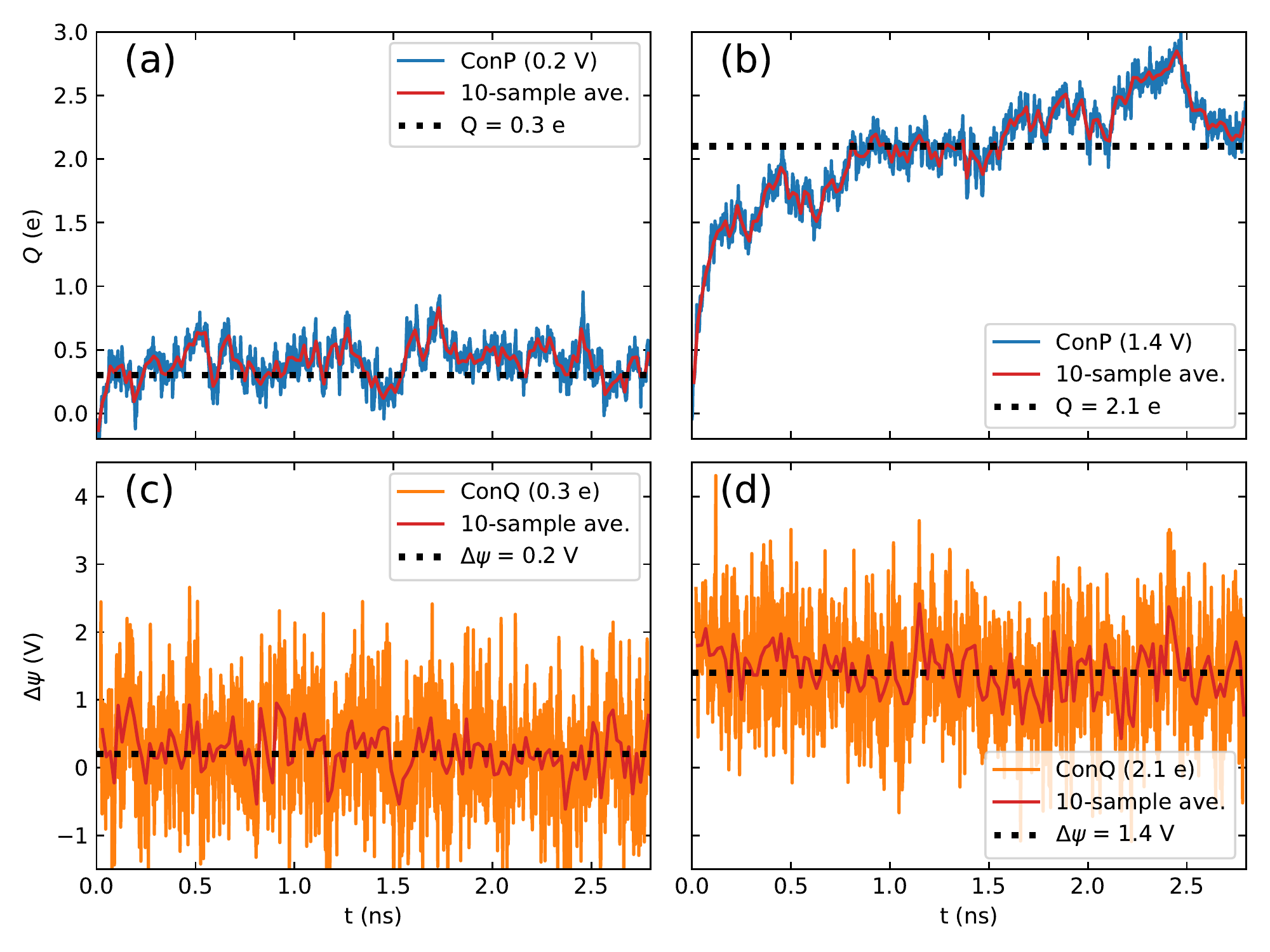}
    \caption{
    Trajectories of (top) electrode charge in ConP simulations and (bottom) potential difference in ConQ simulations for an ionic liquid supercapacitor for the parameters (left) $\Delta \psi = 0.2$ V or $Q = 0.3$ e, or (right) $\Delta \psi = 1.4$ V or $Q = 2.1$ e.
    After initial equilibration, ConP and ConQ simulations start to show ensemble equivalence of expectation values, as indicated by the dotted lines.
    The red traces are averages taken every ten samples (with one sample taken every 2 ps) and highlight how fluctuation magnitudes and time scales differ between ConP and ConQ simulations.
    ConP simulations have relatively smaller fluctuations in the electrode charge, and a significant portion of the variation occurs on multi-nanosecond time scales.
    By contrast, ConQ simulations have relatively larger fluctuations in the potential difference, but most of the variation occurs on picosecond instead of nanosecond time scales.
    }
    \label{fig:timetrace}
\end{figure}

\begin{figure}
    \centering
    \includegraphics[width=\columnwidth]{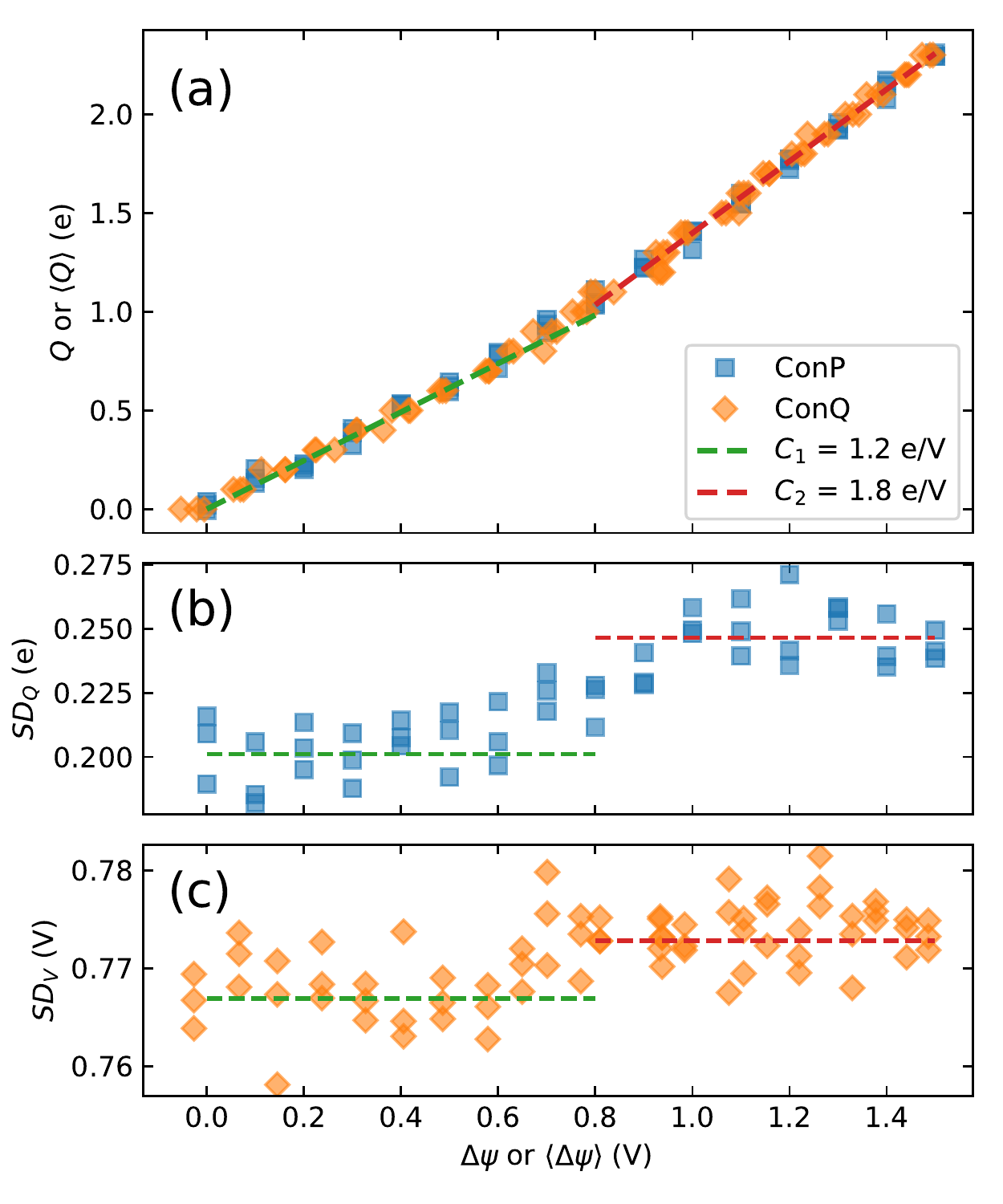}
    \caption{Trajectory averages (a) and standard deviations (b, c) for charge in ConP simulations or potential difference in ConQ simulations. (a) Within statistical uncertainties, ConP and ConQ simulations are thermodynamically equivalent based on the averaged charge and potential differences. Both ensembles give equivalent capacitances based on charge-potential difference curves. We broadly divide the data into two regions of capacitance $C_1 = $ 1.2 e/V and $C_2 = $ 1.8 e/V respectively, with linear fitting indicated by the dashed lines. (b) The standard deviations of charge in the ConP ensemble are consistent with the fitted capacitances, with a clear jump at $\Delta \psi \approx$ 0.8 V. (c) The standard deviations of potential difference in the ConQ ensemble are also consistent with the fitted capacitances, but the difference in standard deviations is less statistically significant relative to noise.}
    \label{fig:sigma_SDs}
\end{figure}

In Figure \ref{fig:timetrace}, typical trajectories of electrode charge in ConP simulations are compared with typical trajectories of potential difference in ConQ simulations.
These trajectories demonstrate the thermodynamic equivalence of equilibrium expectation values.
For example, in ConP simulations performed at $\Delta \psi$ = 0.2 V, the electrodes evolve an average charge of ($\pm$) 0.3 e;
in the corresponding ConQ simulations performed at $Q = 0.3$ e, the electrodes evolve an average potential difference of 0.2 V.
The thermodynamic equivalence is seen to hold across the complete range of potential differences and electrode charges used, as depicted in Figure \ref{fig:sigma_SDs}(a).
The capacitance curves of electrode charge against potential difference are statistically equivalent whether calculated from ConP or ConQ simulations.
Since the electrode responses are thermodynamically equivalent for this set ConP and ConQ simulations, the expectation values of electrolyte observables are also equivalent.
For example, the density profiles of anions and cations across the capacitor are indistinguishable when compared for equivalent ConP and ConQ simulations (and thus not shown; see reference \citenum{Tee2022} for typical profiles).

However, the fluctuations in ConP and ConQ simulations have markedly differing characteristics.
In ConP simulations, the fluctuations of electrode charge are relatively small but persist over nanosecond-long time scales.
As Figure 2\ref{fig:timetrace}(a) and (b) show, when moving from sampling every 2 picoseconds to averaging over 20 picoseconds, most of the electrode charge fluctuations persist in ConP simulations.
By contrast, in ConQ simulations, most of the fluctuations in potential difference occur over picosecond time scales, and averaging over just 20 picoseconds removes significant variability compared to sampling every 2 picoseconds.

In principle, the standard deviations of potential difference (in ConQ simulations) or electrode charge (in ConP simulations) are related to the differential capacitance of the simulated supercapacitor, through equations \eqref{eq:varconq} and \eqref{eq:varconp} respectively.
Figure \ref{fig:sigma_SDs} shows that, in practice, the standard deviation is more useful in ConP simulations than ConQ simulations.
As Figure \ref{fig:sigma_SDs}(a) shows, the differential capacitance of the simulated supercapacitor is non-linear with respect to the applied potential difference, allowing us to check if the changing capacitance is reflected in observed standard deviations.
In lieu of more precise analysis we divide the observed charge-potential difference curve into two regions: at lower potential differences the capacitance is about 1.2 electrons per volt, while at higher potential difference the capacitance is about 1.8 electrons per volt.
When we calculate the standard deviations of electrode charge in ConP simulations (Figure \ref{fig:sigma_SDs} (b)), or of potential difference in ConQ simulations (Figure \ref{fig:sigma_SDs}(c)), these two regions can also be clearly seen.
However, the difference in standard deviations is much clearer for ConP simulations than for ConQ simulations, in line with equations \eqref{eq:varconq} and \eqref{eq:varconp}.

We can further compare the observed standard deviations with the vacuum capacitance, which would have been observed with no electrolyte.
In our simulation this value is $C_0 = 0.056$ e/V (within 5\% of the theoretical value using the parallel plane formula, $C = \epsilon_0 A/d$).
This corresponds to a ConP charge standard deviation of $\sqrt{k_B T C_0} = 0.044$ e, and a ConQ potential difference standard deviation of $\sqrt{k_B T / C_0} = 0.785$ V.
Comparing these values to the standard deviations in Figure \ref{fig:sigma_SDs} shows that in ConP simulations, the standard deviation is much larger than the vacuum capacitance contribution.
In ConQ simulations, however, the observed potential difference standard deviations are much closer to the vacuum capacitance value.
Thus, it is statistically less efficient to use the variance of potential difference from ConQ simulations for calculating the differential capacitance, compared to using the variance of electrode charge from ConP simulations.
Note that this is primarily a relative inefficiency -- even in ConP simulations, standard deviations typically vary by 10\% between runs (and thus variances by 20\%), limiting the accuracy of any calculated capacitance.

\subsection{Configurational Variance from Fluctuations of Electrolyte-Induced Charge}

\begin{figure}
    \centering
    \includegraphics[width=\columnwidth]{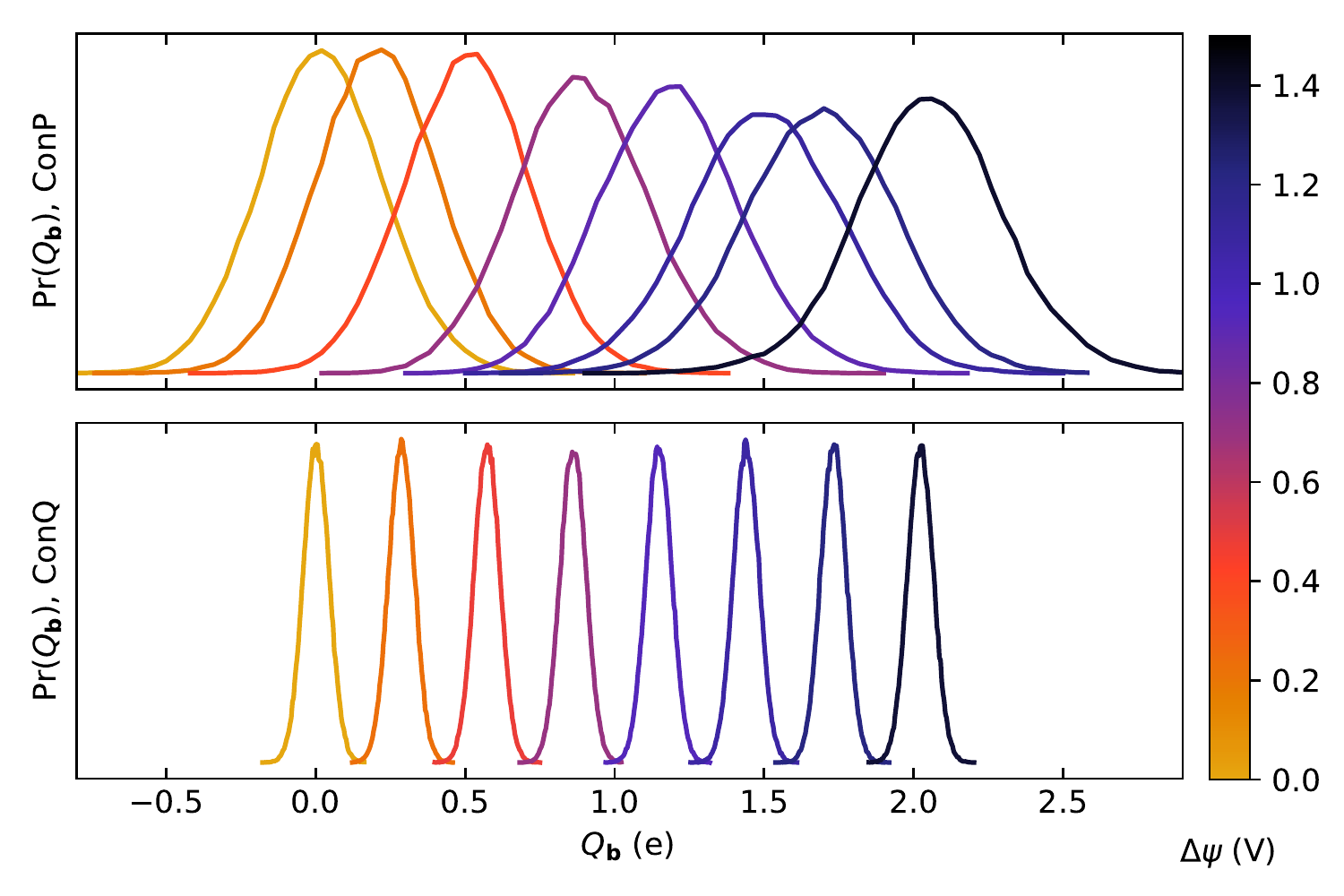}
    \caption{Histograms of electrolyte-induced charge, $Q_\mathbf{b}$ in ConP (top) and ConQ (bottom) for various potential differences (color-coded). At the same potential difference, ConP and ConQ charge distributions show similar means, but the ConP distributions have larger variance and more visible non-normality than the ConQ distributions.}
    \label{fig:sm_histos}
\end{figure}

Figure \ref{fig:timetrace} shows how ConP simulations at distant potential differences have little overlap of electrode charges, while ConQ simulations with a comparable difference in electrode charges have overlapping potential differences.
However, ConQ simulations in fact have a much smaller configurational variance than ConP simulations.
To demonstrate this, we show histograms of the electrolyte-induced charge, $Q_\mathbf{b}$, in ConP and ConQ simulations at various potential differences (Figure \ref{fig:sm_histos}).
Eight potential differences evenly spaced from 0.0 V to 1.6 V were chosen for the ConP simulations for the top graph, and the ConQ simulation with corresponding average potential difference was chosen for the bottom graph.

As Figure \ref{fig:sm_histos} shows, each ConP simulation has a similar mean $Q_\mathbf{b}$ compared to its corresponding ConQ simulation, again demonstrating the thermodynamic equivalence of ensembles.
However, the variances and shapes of the $Q_\mathbf{b}$ distributions clearly differ.
The ConP distributions have far larger variances than corresponding ConQ distributions, and non-normality of the ConP distributions can clearly be seen especially at potential differences of 0.8 V and 1.0 V.
Over that range of potential differences, earlier studies have shown that a transition between different electrolyte packings, stabilized by counter-charges on the electrodes \cite{Merlet2014EDLLife}.
The coexistence of different packings leads to larger configurational variance in the ConP ensemble and correspondingly higher differential capacitances.

Importantly, neighbouring $Q_\mathbf{b}$ distributions significantly overlap in the ConP ensemble, but show negligible overlap in the ConQ ensemble.
Thus, ConP simulations lend themselves to histogram-reweighting analysis \cite{Limmer2013}, where electrode charge probability distributions at one potential difference can be inferred from results at another potential difference.
Multiple simulations can thus be combined into a continuous scan of the electrodes' capacitive response.
With far less overlap between configurational distributions, a histogram-reweighting method to combine multiple ConQ simulations will likely be inefficient compared to using ConP simulations.
Nonetheless, the confinement of ConQ simulations to a narrower distribution of configurations results in significantly faster equilibration, as shown in Figure \ref{fig:timetrace}.
The configurational confinement also results in reduced long-time correlations, as we explore in the next subsection.

\subsection{Correlation Times in ConP and ConQ simulations via Block Standard Error Analysis}

We quantify long-time correlations in ConP and ConQ simulations using block analysis of standard errors \cite{GROSSFIELD200923}.
In block analysis, a trajectory of duration $N \tau$ is divided into $N$ blocks, each of duration $\tau$.
An observable $x$ is averaged over each of the $N$ blocks to give $N$ block averages $x_1, x_2, \ldots, x_N$.
The block standard error (BSE) estimate is then the standard deviation of the $N$ block averages divided by the square root of $N$:
\begin{equation}
    \mathrm{BSE}(\tau) \equiv \frac{\sqrt{\sum_{i=1}^{N}(x_i - \overline{x})^2/(N-1)}}{\sqrt{N}}
\end{equation}
If some time scale $\tau_0$ is large enough that the $N$ block averages are independent and identically distributed, then (by the Central Limit Theorem) the estimated standard error should be constant for all longer time scales $\tau > \tau_0$.
On the other hand, if there are significant correlations outlasting $\tau_0$, then the estimated standard error will be smaller at $\tau_0$ than at some longer time scale $\tau > \tau_0$, indicating the correlation of successive block averages.
We use this method instead of calculating the autocorrelation time to avoid the assumption of a single correlation time scale, which is particularly acute in ConQ simulations given the intense and rapid fluctuations in potential difference.

\begin{figure}
    \centering
    \includegraphics[width=\columnwidth]{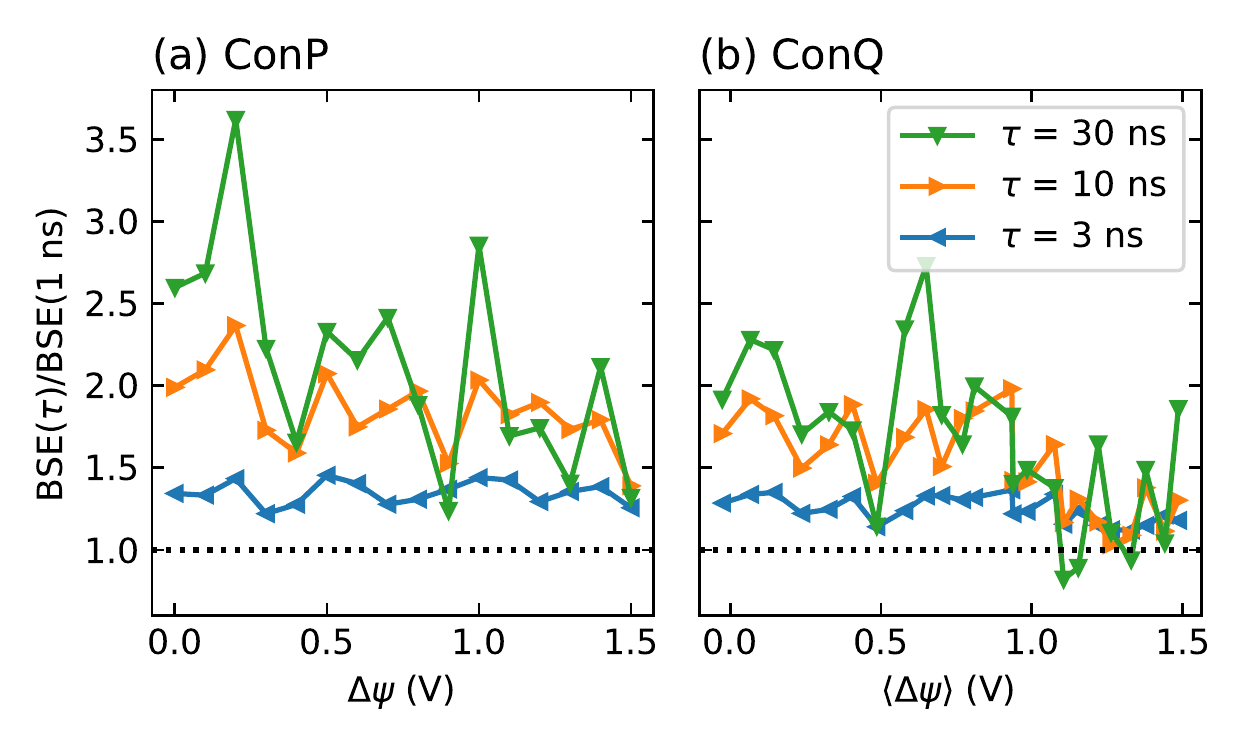}
    \caption{BSE estimate ratios for time scales of 3 ns, 10 ns and 30 ns in ConP (left) and ConQ (right) simulations, compared to a baseline time scale of 1 ns. For ConP simulations, at most potential differences the estimated BSE increases with longer time scale durations, indicating residual long-time correlations at those time scales. By contrast, in ConQ simulations, the estimated BSE increases less with longer time scale durations, and between potential differences of 0.8 to 1.5 V there is particularly little correlation at time scales exceeding 3 ns.}
    \label{fig:blockaves}
\end{figure}

Figure \ref{fig:blockaves} shows the BSE estimated at 3 ns, 10 ns, and 30 ns block durations, relative to the 1 ns BSE, with the three trajectories from different initial conditions combined into one long trajectory for each analysis.
For most ConP simulations, the BSE estimates increase steadily as the block duration increases from 1 ns to 30 ns, indicating that the charges are still correlated at a 10 ns time scale.
Long-time correlations appear to be less prominent at higher potential differences, but there is no clear systematic trend.
By contrast, most of the ConQ simulations show less correlation on the same time scales, since the increases in BSE estimates are smaller compared to the increases for ConP simulations.
In particular, the BSE estimates do not increase much going from block durations of 10 ns to 30 ns, indicating that most of the correlation in ConQ simulations occurs over the 1 to 10 ns time scale.
Indeed, for imposed charges larger than 1.0 electrons per electrode (corresponding to observed potential differences between 0.9 and 1.5 V), the BSE estimates hardly increase over all block durations, indicating minimal correlations exceeding the nanosecond time scale.
Over this region, the ConQ simulations exhibit faster equilibration dynamics compared to ConP simulations at the same potential difference.

As previously mentioned, the coexistence of different packing arrangements of the electrolyte lies behind the increased differential capacitance between 0.9 and 1.5 V (Figure \ref{fig:sigma_SDs}) \cite{Merlet2014EDLLife}.
Intuitively, a ConP simulation has the configurational freedom to sample these different packing arrangements, leading to slow dynamics associated with transitions from one arrangement to another.
On the other hand, a ConQ simulation is more configurationally confined to the most favourable packing arrangement at the specified electrode charge, minimizing the effect of transitional dynamics and thus reducing long-time correlations.

\section{Discussion and Conclusions}

In this study, we have shown how the ConQ ensemble for electrode-electrolyte simulations is theoretically interesting and practically valuable.
Our statistical-mechanical analysis shows how ConQ simulations are thermodynamically equivalent to the more common ConP simulations, validating increasing interest in these simulations, and presents a simple physical picture to underlie these findings.
While ConQ and ConP simulations should return identical ensemble averages (with sufficient sampling and system size), the configurational fluctuations in each ensemble are different.
ConQ simulations fix an extensive variable -- the total electrode charges -- and thus show narrow fluctuations and less configurational overlap between simulations at different phase points.
On the other hand, ConP simulations fix an intensive variable -- the electrode potential differences -- and thus show wider fluctuations, with the configurational overlap making histogram reweighting and other multi-replica methods more feasible.

We demonstrate these features in ConP and ConQ simulations of a typical ionic liquid-graphene supercapacitor.
Both ConP and ConQ simulations give statistically identical plots of charge against potential difference.
Standard deviations of relevant observables (electrode charge in ConP simulations and potential difference in ConQ simulations) are consistent with the changing capacitance at different potential differences, as theory predicts.
ConQ simulations also show less configurational fluctuation as expected, which leads both to less configurational overlap between simulations and smaller long-time correlations.

Given our results, we expect that ConQ simulations will become an important tool in the computational toolbox for studying electrode-electrolyte interfaces, for the following reasons.
Firstly, the constant total electrode charge condition is congruent with quantum-mechanical methods, making direct comparison between ConQ and DFT simulation results more amenable than comparisons between ConP simulations and grand-canonical DFT calculations.
Secondly, the control of an extensive variable in ConQ simulations seems to favour tight sampling of the most favourable electrolyte configuration at a given electrode charge.
This may yield better insights into the electrolyte configurational response as a function of electrode charge with better computational efficiency.
Finally, as other emerging papers in the field indicate, ConQ simulation methods with time-varying electrode charges are analogues \emph{in silico} to amperometric characterization, which is a key technique for electrochemical characterization in the laboratory.
Taken together, these reasons point to ConQ simulation as an emerging technique which is likely to yield new insights into the structure and dynamics of electrochemical interfaces.

\begin{acknowledgement}

The authors thank the Australian Research Council for its support for this project through the Discovery program (FL190100080). We thank Prof. Robert Mei{\ss}ner and Ludwig Ahrens-Iwers for their invaluable support in developing and debugging the source code used in this project. We acknowledge access to computational resources at the NCI National Facility through the National Computational Merit Allocation Scheme supported by the Australian Government, and this work was also supported by resources provided by the Pawsey Supercomputing Centre with funding from the Australian Government and the government of Western Australia. We also acknowledge support from the Queensland Cyber Infrastructure Foundation (QCIF) and the University of Queensland Research Computing Centre (RCC).

\end{acknowledgement}

\begin{suppinfo}

\section{ConQ Thermodynamic Ensembles: Full and Born-Oppenheimer}

In constant potential molecular dynamics simualtions we generally use the Born-Oppenheimer approach, where the fast degrees of freedom (in this case in the conductive charge redistribution) is instantaneously energy-minimized with respect to the slow degrees of freedom that are time-integrated (in this case the electrolyte configuration).
Scalfi et al. \cite{Scalfi2020ChargeEnsemble} show that in the case of ConP simulations, where the potential difference is held constant, the Born-Oppenheimer ensemble loses degrees of freedom relative to a ``full'' ensemble (in which the fast degrees of freedom would also be explicitly integrated and have their own thermal variation).
Notably, the differential capacitance calculated as the variance of a Born-Oppenheimer ensemble simulation is less than the actual differential capacitance (from the observed charge-potential difference curve), the discrepancy corresponding to the vacuum capacitance.

In this Appendix we use similar methods to develop the Born-Oppenheimer and full ConQ ensembles.
These provide mathematical justification for the expression in \eqref{eqn:soln_qdagger} for $\mathbf{q}_Q$, and shed light on the Legendre transforms between ConP and ConQ ensembles.
We reuse Scalfi et al.'s notation as much as possible, but the derivations here are self-contained.
As such, we will repeat key results from that paper without proof where appropriate.
As a minor difference, we write delta functions of the charge rather than charge per thermal energy -- that is, $\delta(Q)$ rather than $\delta(\beta Q)$ -- to give proper units in the partition functions.

\subsection{The Full ConQ partition function}
\label{sec:SI_fullzq}

In the full ConQ partition, the electrolyte configuration $\mathbf{r}^{3N}$ and electrode charge $\mathbf{q}$ can vary independently.
(In the subsequent Born-Oppenheimer approach, the latter will be a function of the former.)
Then the probability of a joint configuration $(\mathbf{r}^{3N}, \mathbf{q})$, with the total electrode charge constrained to $Q$, is proportional to:
\[
  \mathrm{Pr}_Q(\mathbf{r}^{3N}, \mathbf{q}) \propto \exp \left[- \beta \left( U_0(\mathbf{r}^{3N}) + \frac{1}{2} \mathbf{q}^T \mathbf{A} \mathbf{q} - \mathbf{b}^T(\mathbf{r}^{3N}) \mathbf{q} \right) \right] \delta (\mathbf{e}^T \mathbf{q}) \delta (\mathbf{d}^T \mathbf{q} - Q)
\]
with notation defined as in the main text, equations \eqref{eqn:def_U} and \eqref{eqn:def_d}.
The Boltzmann term is proportional to the total system energy -- note no potential difference is imposed across the electrodes -- while the two Dirac delta functions enforce charge constraints.
The first delta function enforces electroneutrality, keeping the total charge zero:
\[
 \mathbf{e}^T \mathbf{q} \equiv \{1, \cdots, 1\} \cdot \mathbf{q} = 0
\]
while the second delta function enforces the electrode charge constraint.

Now, using the Fourier representations:
\[
    \delta(\e^T \q) = \frac{1}{\beta} \delta(\beta \e^T \q) = \frac{1}{2\pi \beta} \idke \ e^{i k_e \beta \e^T \q}; \, \, \, \, \,  \delta(\beta \dd^T \q) = \frac{1}{2\pi \beta} \idkd \ e^{i k_d \beta (\dd^T \q - Q)}
\]
we can write the full ConQ configurational partition function as:
\begin{align}
  Z_{\mathrm{full}}(Q) &= \idrn \ebuO \idq \exp \left[-\beta \left(\frac{1}{2} \q^T \A \q - \bb^T \q\right) \right] \delta(\e^T \q) \delta(\dd^T \q - Q)
  \label{eqn:SI_zfullQ}\\
  &= \frac{1}{(2 \pi \beta)^2}  \idrn \ebuO \idke \idkd G_{\q,Q}(k_e, k_d); \nonumber \\
  G_{\q,Q}(k_e, k_d) 
  &\equiv \idq \exp \left[-\frac{1}{2} \beta \q^T \A \q + \beta \left( \bb^T + ik_e \e^T + ik_d \dd^T \right)\q - i \beta k_d Q\right] \nonumber
\end{align}
where we isolate $G_\q$, the Gaussian integral in $\q$.
We complete the square in $\q$ and use Gaussian identities to evaluate $G_\q$, writing $\C \equiv \A^{-1}$ for convenience (as in the main text), and remembering that $M$ is the number of electrode particles, or the dimensionality of $\q$:
\begin{align*}
G_{\q,Q}(k_e, k_d)&= \sqrt{\frac{(2 \pi)^M \det \C}{\beta^M }} \exp \left[\frac{\beta}{2} \left(\bb + i k_e \e + i k_d \dd \right)^T \C \left(\bb + i k_e \e + i k_d \dd \right) -i\beta k_d Q\right] \\
    &= \sqrt{\frac{(2 \pi)^M\det \C}{\beta^M }} \exp \left[\beta \left(\frac{1}{2} \bb^T \C \bb - \frac{1}{2} \e^T \C \e \,\, k_e^2 + \e^T \C(i \bb - k_d \dd)\,\, k_e \right. \right. \\ & \left. \left. - \frac{1}{2} \dd^T \C \dd \, \, k_d^2 + i (\dd^T \C \bb - Q) \,\, k_d \right) \right]
\end{align*}
%

We can then complete the square in $k_e$ to perform the next Gaussian integral:
\begin{align*}
    \idke G_{\mathbf{q},Q}(k_e, k_d) &= \sqrt{\frac{(2 \pi)^M \det \C}{\beta^M}} \exp \left[\beta \left( \frac{1}{2} \bb^T \C \bb - \frac{1}{2} \dd^T \C \dd \, \, k_d^2 + i (\dd^T \C \bb - Q) \,\, k_d \right) \right] \\ 
    &\hspace{2em} \times \sqrt{\frac{2 \pi}{\beta \e^T \C\e}} \, \exp \left[\frac{\beta}{2 \e^T \C \e} (i \bb - k_d \dd)^T \C^T \e \e^T \C  (i \bb - k_d \dd) \right] \\
    &= \sqrt{\frac{(2 \pi)^M \det \C}{\beta^M }} \sqrt{\frac{2 \pi}{\beta \e^T \C\e}} \exp \left[\frac{\beta}{2} \bb^T \OO \C \bb -\frac{\beta}{2} \dd^T \OO \C \dd\,\, k_d^2 + i \beta (\dd^T \OO \C \bb - Q)\,\, k_d \right].
\end{align*}
Here we have defined the electroneutrality matrix (recalling that $\C^T = \C$)
\[
\OO \equiv \mathbf{I} - \frac{\C \e \e^T}{\e^T \C \e}
\]
which ensures the electroneutrality of the final charge configuration. Finally, completing the square in $k_d$:
\begin{align*}
\idkd \idke G_{\mathbf{q},Q}(k_e, k_d) &= \sqrt{\frac{(2\pi)^M \det \C}{\beta^M}}\sqrt{\frac{2\pi}{\beta \e^T \C \e}}\sqrt{\frac{2\pi}{\beta \dd^T \OO \C \dd}} \\ &\hspace{2em}\exp \left[\frac{\beta}{2} \left(\bb^T \OO \C \bb - \frac{(Q - \dd^T \OO \C \bb)^2}{\dd^T \OO \C \dd} \right) \right] \\
& \equiv \sqrt{\frac{(2\pi)^M \det \C}{\beta^M}}\sqrt{\frac{2\pi}{\beta \e^T \C \e}}\sqrt{\frac{2\pi}{\beta C_0}} \exp \left[- \beta \left(- U_\bb +\frac{1}{2} C_0 \Delta \psi_Q (Q, \bb)^2 \right) \right]
\end{align*}
where in the second line we reuse the following definitions from the text:
\begin{itemize}
    \item vacuum capacitance, $C_0 \equiv \dd^T \OO \C \dd$
    \item ConQ electrolyte charge $Q_\bb \equiv \dd^T \OO \C \bb$ and potential difference $\Delta \psi_Q \equiv (Q - Q_\bb)/C_0$
    \item electrolyte polarization energy $U_\bb \equiv (\bb^T \OO \C \bb)/2$
\end{itemize}

In summary, the Gaussian integral over $G_{\q,Q} (k_e, k_d)$ has evaluated to several pre-factors, associated with the integrated-over degrees of freedom, multiplied by the Boltzmann factor of the electrode energy $U_Q$ (equation \eqref{eqn:def_U}) with Born-Oppenheimer electrode charges
\[ \q_Q (Q, \bb) \equiv \OO \C(\bb + \Delta \psi_Q(Q, \bb)). \]
We can thus write a compact expression for the full ConQ partition function, in equation \eqref{eqn:SI_zfullQ}, as:
\begin{equation}
  Z_{\mathrm{full}}(Q) = \sqrt{\left(\frac{2 \pi}{\beta} \right)^{M-1}  \frac{\det \C}{\e^T \C \e} } \sqrt{\frac{\beta}{2 \pi C_0}} \idrn \exp \left[- \beta \left(U_0(\rN) - U_\bb +\frac{1}{2} C_0 \Delta \psi_Q (Q, \bb)^2 \right) \right]
  \label{eqn:SI_fullqz_expression}
\end{equation}

\subsection{Laplace transforms between ConP and ConQ partition functions}
\label{sec:SI_comparisons}

We can now list expressions for the full and Born-Oppenheimer partition functions for the ConQ ensembles (not numbering the first equation which repeats equation \eqref{eqn:SI_zfullQ}), which are original to this paper.
We also show partition functions for the ConP ensembles, which were derived in Scalfi et al.\cite{Scalfi2020ChargeEnsemble} in similar fashion:
\begin{align}
  Z_{\mathrm{full}}(Q) &\equiv \idrn \ebuO \idq \exp \left[-\beta \left(\frac{1}{2} \q^T \A \q - \bb^T \q\right) \right] \delta(\e^T \q) \delta(\dd^T \q - Q)
  \nonumber \\
  Z_{\mathrm{BO}}(Q) &\equiv \idrn \ebuO \exp \left[-\beta \left(- U_\bb + \frac{1}{2} C_0 \Delta \psi_Q (Q, \bb)^2 \right) \right] \equiv Z_Q(Q)
  \label{eqn:SI_zOBQ} \\
  \Rightarrow Z_{\mathrm{full}}(Q) &= \sqrt{\left(\frac{2 \pi}{\beta} \right)^{M-1}  \frac{\det \C}{\e^T \C \e} } \sqrt{\frac{\beta}{2 \pi C_0}} \times Z_{\mathrm{BO}}(Q);
  \label{eqn:SI_zOBQ_relationship} \\
  Z_{\mathrm{full}}(\Delta \psi) &\equiv \idrn \ebuO \idq \exp \left[-\beta \left(\frac{1}{2} \q^T \A \q - \bb^T \q - \Delta \psi \dd^T \q \right) \right] \delta( \e^T \q)
  \label{eqn:SI_zfullV}\\
 Z_{\mathrm{BO}}(\Delta \psi) &\equiv \idrn \ebuO \exp \left[-\beta \left(- U_\bb + \frac{1}{2} C_0 \Delta \psi^2 - \Delta \psi Q_P(\Delta \psi, \bb) \right) \right] \equiv Z_P(\Delta \psi)
  \label{eqn:SI_zOBV} \\
  \Rightarrow Z_{\mathrm{full}}(\Delta \psi) &= \sqrt{\left(\frac{2 \pi}{\beta} \right)^{M-1}  \frac{\det \C}{\e^T \C \e} } \times Z_{\mathrm{BO}}(\Delta \psi).
  \label{eqn:SI_zOBV_relationship}
\end{align}

Relations \eqref{eqn:SI_zOBQ_relationship} and \eqref{eqn:SI_zOBV_relationship} result clearly from the reduced degrees of freedom associated with the Born-Oppenheimer procedure.
The full ConP ensemble is constrained to a  $3N + M - 1$ dimensional hypersurface due to the $N$ electrolyte positions and $M$ electrode charges, minus one for the electroneutrality constraint.
Meanwhile the full ConQ ensemble is constrained to a $3N + M - 2$ dimensional hypersurface due to the additional constraint fixing the total electrode charge.
However, both Born-Oppenheimer partition functions are integrated over only $3N$ degrees of freedom, because the explicit dependence of $\mathbf{q}$ on $\mathbf{r}^{3N}$ removes $M-1$ degrees of freedom in the ConP ensemble and $M-2$ in the ConQ ensemble.

Thus, while an exact Laplace transform can hold between the full ConP and ConQ partition functions, the Laplace transform between the Born-Oppenheimer versions has a scaling factor as described in the main text.
The exact transform between full partition functions is easily exhibited by inspection. 
We can simply substitute $Z_{\mathrm{full}}(Q)$, as defined in \eqref{eqn:SI_zfullQ}, into the transform equation
\[
  Z_{\mathrm{full}}(\Delta \psi) = \int_{-\infty}^{\infty} \mathrm{d}Q \,e^{\beta \Delta \psi Q} Z_{\mathrm{full}}(Q)
\]
and verify that $Z_{\mathrm{full}}(\Delta \psi)$ as defined \eqref{eqn:SI_zfullV} is obtained as the Legendre transform.
Then the Laplace transform between Born-Oppenheimer partition functions is easily shown using equations \eqref{eqn:SI_zOBQ_relationship} and \eqref{eqn:SI_zOBV_relationship}, completing the derivation of equation \eqref{eqn:soln_legendre} in the main text.

\end{suppinfo}

\bibliography{achemso-demo}

\end{document}